\documentclass[a4paper,11pt]{article}
\pdfoutput=1 
\usepackage{jheppub} 
\usepackage{graphicx,array}
\usepackage{makecell}
\usepackage{hyperref}
\usepackage{xcolor}
\usepackage{amsmath,amssymb,slashed,latexsym}
\usepackage{bm}
\usepackage{braket}
\usepackage{placeins}
\usepackage{adjustbox}
\usepackage[caption=false]{subfig}
\usepackage{enumitem}
\usepackage{amsthm}

\usepackage{textcomp}

\newcommand{\ord}[1]{\mathcal{O}{(#1)}}

\usepackage{array}
\newcolumntype{P}[1]{>{\centering\arraybackslash}p{#1}}
\newcolumntype{M}[1]{>{\centering\arraybackslash}m{#1}}

\definecolor{darkgreen}{rgb}{0.0, 0.8, 0.0} 
\definecolor{darkblue}{cmyk}{1,0.4,0,0.3}
\definecolor{violet}{cmyk}{0,1,0,0.2}
\hypersetup{colorlinks, bookmarksnumbered, citecolor=darkblue, linkcolor=darkblue, pdfstartview=FitH, urlcolor=darkblue, linktocpage}

\newcommand{\GeV}{\mathrm{GeV}}
\newcommand{\TeV}{\mathrm{TeV}}

\def\beq{\begin{equation}}
\def\eeq{\end{equation}}
\def\beqa{\begin{eqnarray}}
\def\eeqa{\end{eqnarray}}

\newcommand{\cO}{\mathcal{O}}
\newcommand{\cL}{\mathcal{L}}
\newcommand{\cP}{\mathcal{P}}

\newcommand{\eg}{\textit{e.g. }}

\newcommand{\fb}{\mathrm{fb}}

\title{\center New Physics at the Muon (Synchrotron) Ion Collider: MuSIC for several scales}

\preprint{MIT-CTP/5811}

\author[1]{Hooman Davoudiasl,}
\author[1,2]{Hongkai Liu,}
\author[3]{Roman Marcarelli,}
\author[2]{Yotam Soreq,}
\author[4]{Sokratis Trifinopoulos}

\affiliation[1]{High Energy Theory Group, Physics Department \\ Brookhaven National Laboratory,
Upton, NY 11973, USA}
\affiliation[2]{Physics Department, Technion – Israel Institute of Technology, Haifa 3200003, Israel}
\affiliation[3]{Department of Physics, University of Colorado, Boulder, Colorado 80309, USA}
\affiliation[4]{Center for Theoretical Physics, Massachusetts Institute of Technology, Cambridge, MA 02139, USA}

\abstract{
A Muon (Synchrotron) Ion Collider~(MuSIC) can be the successor to the Electron-Ion Collider at  Brookhaven National Laboratory, as well as the ideal demonstrator facility for a future multi-TeV Muon Collider. 
Besides its rich nuclear physics and Standard Model particle physics programs, in this work we show that the MuSIC with a TeV-scale muon beam offers also a unique opportunity to probe New Physics. 
In particular, the relevant searches have the potential to surpass current experimental limits and explore new regimes of the parameter space for a variety of Beyond the Standard Model scenarios including: lepton-flavor violating leptoquarks, muonphilic vector boson interactions, axion-like particles coupling to photons, and heavy sterile neutrinos. 
Depending on the particular case, the sensitivity of the searches in the MuSIC may span a wide range of energy scales, namely from sub-GeV particles to the few TeV New Physics mediators. Our analysis demonstrates  that the MuSIC can strike a powerful chord in the search for New Physics, thanks to unique combination of features that amplify its capabilities.
}

\begin{document}

\maketitle

\flushbottom

\section{Introduction}
\label{sec:intro}

The development of new experimental facilities is crucial for advancing our understanding of the fundamental constituents of matter and their interactions. 
As we approach the High-Luminosity phase of the LHC~(HL-LHC), several new proposals are being considered for the post-LHC era, including 
the electron-positron and hadron-hadron Future Circular Collider (FCC-ee~\cite{FCC:2018byv,FCC:2018evy} and FCC-hh~\cite{FCC:2018vvp}), as well as the multi-TeV Muon Collider (MuC)~\cite{MuonCollider:2022nsa,Accettura:2023ked,InternationalMuonCollider:2024jyv}. 
These and other envisioned future collider facilities are being debated at the forefront of strategic planning in high energy particle physics, with their completion projected over the coming decades. 
Meanwhile, the construction of the first high-intensity Electron-Ion Collider~(EIC)~\cite{Accardi:2012qut,Aschenauer:2014cki} is commencing at Brookhaven National Laboratory~(BNL). As the EIC era is starting to take shape, one could  entertain ideas that look beyond the horizon for what may come next.  
In particular, one could consider how the research and development (R\&D) for future high energy colliders might have natural synergies with the next phase of nuclear physics facilities that come after the EIC.

In this work, motivated by the preceding question, we explore the potential reach of a Muon (Synchrotron) Ion Collider~(MuSIC) for Beyond the Standard Model~(BSM) searches. 
This machine was first conceptualized in Refs.~\cite{Ginzburg:1998yw,Sultansoy:1999na,Acar:2016rde,Canbay:2017rbg,Ketenoglu:2018fai,Kaya:2019ecf,Cheung:2021iev,Dagli:2022idi,Acosta:2021qpx,Acosta:2022ejc,Hatta:2023fqc}, and in Refs.~\cite{Acosta:2021qpx,Acosta:2022ejc}. It was put forward as a realistic successor to the EIC after its mission is completed in the 2040's. The science case with respect to new regimes of deep inelastic scattering for Standard Model~(SM) particle physics and nuclear physics studies was outlined. 
The MuSIC would thus serve as an interesting concept for the next generation of lepton-ion colliders in nuclear physics, while providing a platform for developing a TeV-scale muon synchrotron towards a multi-TeV MuC for high energy physics.  
Here, we also note that another possible candidate that has been discussed in the literature for the post-EIC era is the Large Hadron electron Collider (LHeC)~\cite{LHeC:2020van} at CERN, which we will refer to  later in this work. 

The MuSIC can take advantage of the existing infrastructure at BNL. 
The electron storage ring would be converted into a synchrotron serving as a (anti-)muon storage ring by upgrading the magnets, and adding shielding to protect against the charged particles from (anti-)muon decays. 
Hazards from collimated neutrino radiation are less relevant at lower beam energies~\cite{InternationalMuonCollider:2024jyv}, and additionally Ref.~\cite{Acosta:2022ejc} has proposed a strategy to mitigate them further. 
Moreover, a high-intensity (anti-)muon source could be generated by the proton driver scheme~\cite{Delahaye:2019omf} (utilizing existing high-intensity proton sources) followed by muon cooling (see Refs.~\cite{Terzani:2024neq,Aritome:2024jiv} for recent developments).  
As noted in Ref.~\cite{Acosta:2022ejc}, a staged development can be implemented. In this work, we consider the ``realistic'' benchmark of Ref.~\cite{Acosta:2021qpx} with center-of-mass energy $\sqrt{s}$ and integrated luminosity $L_I$ (estimated per five years of operation)~\cite{Acosta:2021qpx,Acosta:2022ejc} for the MuSIC muon-proton collisions
\begin{equation}
    \label{eq:sLI}
    \sqrt{s} = 1\,\TeV\, , 
    \qquad \qquad 
    L_I = 400\,\fb^{-1}\, .
\end{equation}
The center-of-mass energy is calculated assuming a muon beam of $1\,\TeV$, which can be considered a milestone towards a $3\,\rm TeV$ MuC, and a proton beam of $275\,\GeV$. In case of an ion beam the energy is set to be $100\,\GeV$ per nucleon, and $L_I$ is taken to be $400\,\fb^{-1} / A$, where $A$ is the atomic mass number of the ion. As in the EIC, the ion beam could consist of a large variety of ions, from protons to uranium~\cite{Accardi:2012qut,Aschenauer:2014cki}. 

Similarly to the EIC, besides its rich prospects for the study of SM particle and nuclear physics, the MuSIC also offers a compelling BSM program (\eg see Refs.~\cite{Gonderinger:2010yn,Liu:2021lan,Cirigliano:2021img,Davoudiasl:2021mjy,Yan:2021htf,Li:2021uww,Batell:2022ogj,Zhang:2022zuz,Yan:2022npz,Boughezal:2022pmb,Davoudiasl:2023pkq,Balkin:2023gya,Davoudiasl:2024vje,Wang:2024zns,Wen:2024cfu} for BSM studies at the EIC). 
First of all, having a rather clean initial state on the lepton beam side (up to electroweak corrections) reduces QCD backgrounds and pileup contamination. 
On the ion beam side, \emph{coherent} scattering induced by virtual photons leads to cross sections enhanced by the square of $Z$, the atomic number of the ions. 
Moreover, the MuSIC would inherit the EIC's excellent detection capabilities, especially in the high-rapidity regime~\cite{Adkins:2022jfp,Jentsch}.

Altogether, the MuSIC offers a unique combination of features that we would like to highlight in this work. 
We plan to do this by direct comparison with other proposed future facilities with respect to BSM scenarios which are known to lie within the experimental reach of the respective facilities. 
In particular, we will compare with the following: i) the HL-LHC with $3\,\rm ab^{-1}$ integrated luminosity, ii) Muon-Beam-Dump experiments, which are one of the proposals for the exploration of New Physics during the R\&D phase of the MuC program~\cite{Cesarotti:2022ttv,Cesarotti:2023sje,Sieber:2023nkq,Batell:2024cdl,Fayet:2024ddk} iii) FCC-ee and LHeC, which are other candidates for the next generation of lepton and lepton-ion colliders, respectively.
For concreteness, we study simplified models with specific New Physics states, which are well-motivated from theory. Moreover, in this work, we assume a muon beam, but the prospects of the New Physics cases that we consider would remain basically unchanged in case an anti-muon beam is utilized.  This option may be facilitated by new cooling techniques applicable to a $\mu^+$ beam \cite{Aritome:2024jiv} and has been invoked in relation to proposals for new collider configurations \cite{Hamada:2022mua}.

The paper is structured around the searches for four different New Physics scenarios: lepton-flavor violating leptoquarks in Section \ref{sec:LQs}, muonphilic $Z'$ vector bosons in Section \ref{sec:muonphilic_vectors}, axion-like particles in Section \ref{sec:ALPs}, and the active to sterile transition neutrino magnetic moment in Section \ref{sec:heavy_neutrinos}. 
In each of these sections, we first present the BSM models by writing the simplified Lagrangians. 
Subsequently, we provide our results as exclusion limits in the two-dimensional planes of the relevant New Physics parameter space and compare them with other future experiments. 
Finally, we conclude in Section~\ref{sec:conclusions} and outline our suggestions for future research directions.

\section{Lepton-flavor violating leptoquarks}
\label{sec:LQs}

%
\FloatBarrier
\begin{figure}
 \centering
  \includegraphics[width=0.45\linewidth]{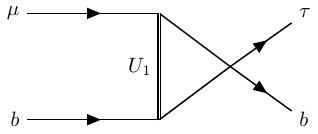}
 \includegraphics[width=0.45\linewidth]{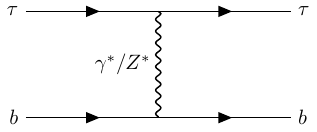}
 \includegraphics[width=0.75\linewidth]{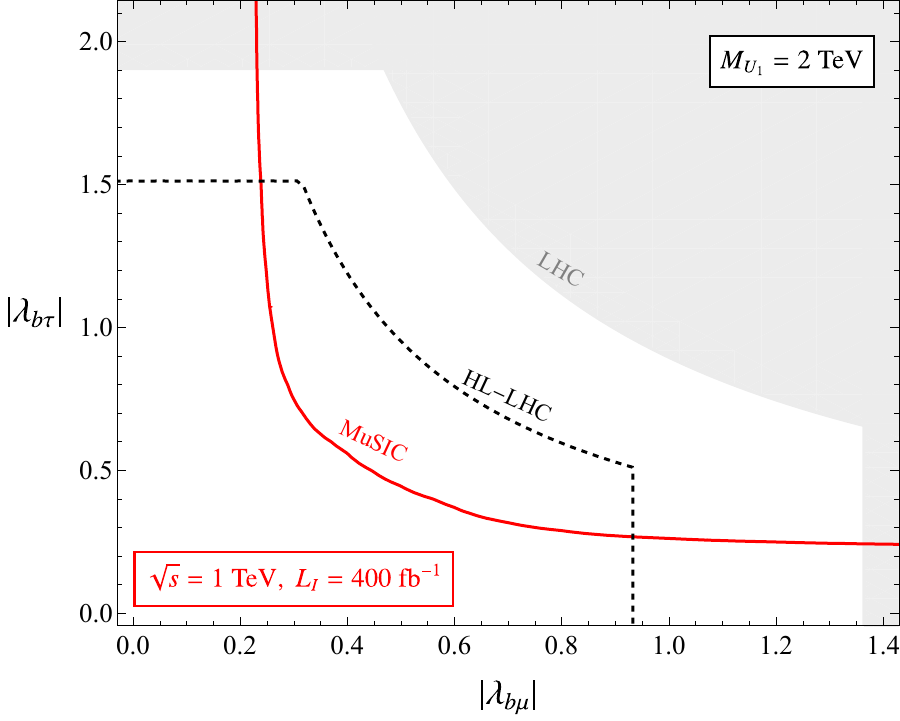}
    \caption{Top: The Feynman diagrams of the LFV process $b\mu\to b\tau$ via $U_1$ leptoquark exchange (left) and the SM background process $b\tau\to b\tau$ relying on the $\tau$ content of the muon beam (right). Bottom: Exclusion limits at 2$\sigma$ for the left-handed lepton-flavor violating couplings of a $U_1$ leptoquark with $m_{U_1} = 2\,\TeV$ between muons or taus and bottom quarks. The MuSIC bound is given by the red curve. The current LHC bounds (gray), as well as the projections for the HL-LHC (dashed black) are derived recasting Refs.~\cite{ATLAS:2020zms,CMS:2021ctt,CMS:2022fsw} with the aid of \textsc{HighPT}~\cite{Allwicher:2022mcg}.}
    \label{fig:LQ_bounds}
\end{figure}

The MuSIC offers an excellent environment for studies of lepton-flavor violation~(LFV) between muons and taus. 
We illustrate this using searches for LFV leptoquark interactions. 
Leptoquarks can generate the partonic process $q \mu \to q \tau$ at tree-level, by coupling to quarks and leptons simultaneously~\cite{Dorsner:2016wpm} and are generically predicted in quark-lepton unification models~\cite{Pati:1974yy}. 
For concreteness, we study the vector leptoquark $U_1 \sim ({\bf 3},{\bf 1},2/3)$, which has recently attracted attention in the context of TeV-scale models of flavor~\cite{Bordone:2017bld,Greljo:2018tuh,Blanke:2018sro}, as well as a portal to dark matter~\cite{Guadagnoli:2020tlx,Baker:2021llj}. 

Regarding the flavor structure, we assume that $U_1$ couples predominantly to the third-generation left-handed quarks and second and third generation left-handed leptons. 
The coupling to bottoms is expected to improve the sensitivity to the LFV final state due to the associated $b$-tagging~\cite{Marzocca:2020ueu}. 
At the renormalizable level, the interaction Lagrangian reads
\begin{align}
    \cL_{U_1}^{\rm int} 
=   \lambda_{b\mu} U_1^\alpha \left( V_{ib} \bar{u}_L^i \gamma_\alpha \nu_\mu 
    + \bar{b}_L \gamma_\alpha \mu_L \right) 
    + \lambda_{b\tau} U_1^\alpha \left( V_{ib} \bar{u}_L^i \gamma_\alpha \nu_\tau + \bar{b}_L \gamma_\alpha \tau_L \right) + \text{h.c.}\,,
\end{align}
where $V$ is the CKM matrix. 
The leptoquark has mass $M_{U_1}$ and decay rate $\Gamma_{U_1} =  \left( |\lambda_{b\mu}|^2 + |\lambda_{b\tau}|^2 \right) M_{U_1} / (12 \pi)$. We remain agnostic to the details of the ultraviolet (UV) completion.

The LFV process $b \mu  \to b \tau $ is generated via a $U_1$ exchange in the $u$-channel, where the $b$ quark originates from a proton beam. 
The relevant tree-level diagrams for the signal and dominant SM background are depicted in the top panel of Fig.~\ref{fig:LQ_bounds}.
Similarly to the situation in high-energy MuCs~\cite{Asadi:2021gah,Azatov:2022itm}, taking into account collinear radiation emitted by splitting of the initial muon state leads to the picture of lepton PDFs derived in Refs.~\cite{Han:2020uid,Han:2021kes,Garosi:2023bvq}. 
For the signature channel, the PDF for the valence parton of the muon beam is from the lepton PDF set \textsc{LePDF}~\cite{Garosi:2023bvq}, and the bottom content of the proton is from the proton PDFs \textsc{MMHT2015qe}~\cite{Harland-Lang:2019pla}. 
The calculation of the main SM background, namely $\tau b \to \tau b$ via a $Z$ boson or photon $t$-channel exchange, requires on the other hand, the PDF of the $\tau$ inside the muon. 

Convolving then the PDFs with the analytical expressions for the differential cross sections of the respective processes and integrating over angular distributions, we obtain the number of events per invariant mass bin. The integration limits for the pseudo-rapidities are determined by the physical region of the central detector~\cite{Adkins:2022jfp}, so we impose a cut $|\eta|\lesssim 4$ on the final state particles. The size of the bins is determined according to the energy resolution of the hadronic calorimeter, denoted as $\sigma_E$. 
Following Ref.~\cite{Acosta:2021qpx}, we use
\begin{equation}
    \label{eq:hCalRes}
    \frac{\sigma_E}{E} 
    = 
    \frac{50\,\%}{\sqrt{E [\GeV]}} \oplus 10\,\% \, ,    
\end{equation}
where $\oplus$ means the two terms are added in quadrature. 
The bottom in the final state jet is identified with $80\,\%$ acceptance efficiency assuming a $b$-tagging algorithm as the one used in ATLAS leptoquark searches into $b\tau$ final state~\cite{ATLAS:2021jyv,ATLAS:2023vxj}, and for the $\tau$ hadronic decays we take the ‘medium’ working point of $75\,\%$ acceptance efficiency. 

We derive $95\%$ C.L. projected exclusion limits on the $\lambda_{b\mu}$ - $\lambda_{b\tau}$ parameter space by observing a modification of the expected SM cross section. The dominant effects originate in the last invariant mass bins, since the center-of-mass energy is below the mass threshold.
We present our results in Fig.~\ref{fig:LQ_bounds} (bottom panel) for a leptoquark of mass $M_{U_1}=2\,\TeV$.
Additionally, one can find the bounds derived in the context of hadron colliders via Drell-Yan processes $pp \to \mu \mu (\tau \tau)$ and the LFV $pp \to \mu \tau$. 
In particular, we use the public code \textsc{HighPT}~\cite{Allwicher:2022mcg} and calculate the current LHC bounds (gray curve) with $139\,\fb^{-1}$~\cite{ATLAS:2020zms,CMS:2021ctt,CMS:2022fsw} as well as the future projections for the $3\,\rm ab^{-1}$ HL-LHC (dashed gray curve).

From Fig.~\ref{fig:LQ_bounds} we infer that for $M_{U_1}=2\,\TeV$ the MuSIC has the potential to surpass the current LHC and future HL-LHC sensitivity to couplings above $\cO(0.2)$.
This is achieved thanks to the valence leptonic partons initiating the tree-level signal process and, despite the fact that HL-LHC is a machine which operates at a center-of-mass energy far above the leptoquark mass and its luminosity will be an order of magnitude higher than that of the MuSIC. 
Considering larger leptoquark masses, the sensitivity of the MuSIC naturally decreases, and we have checked that at $M_{U_1}=4\,\TeV$ the derived bounds match the current LHC ones. We note that even though we have neglected right-handed couplings here, they could also produce contributions to LFV processes. In light of this, it is worth mentioning a unique possibility of New Physics searches at the EIC and MuSIC: one could in principle exploit the helicity information of the polarized initial states (up to electroweak corrections), and distinguish between predominantly left- and right-handed New Physics couplings~\cite{Davoudiasl:2021mjy,Boughezal:2022pmb}

\section{\boldmath Muonphilic $Z'$ vector bosons}
\label{sec:muonphilic_vectors}

%
\begin{figure}
 \centering
  \includegraphics[width=0.45\linewidth]{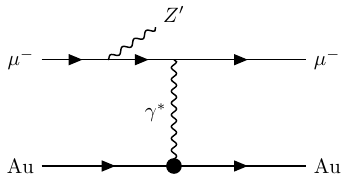}
 \includegraphics[width=0.45\linewidth]{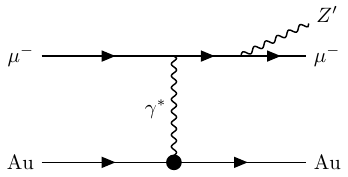}
 \includegraphics[width=0.75\linewidth]{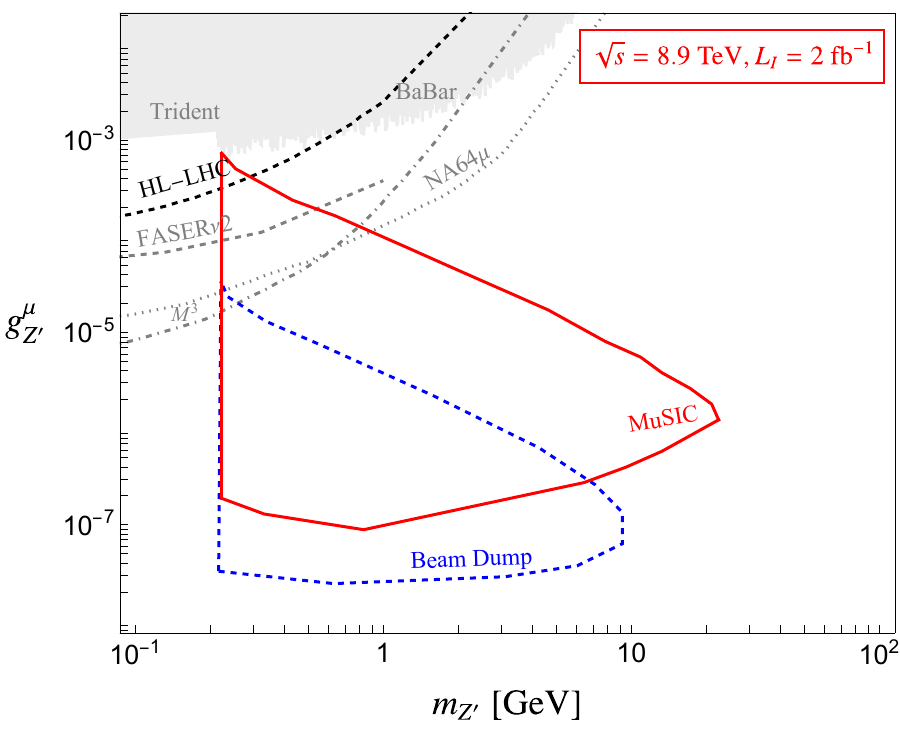}
    \caption{Top: The Feynman diagrams of $Z^\prime$ production at the MuSIC via muon bremsstrahlung. Bottom: Exclusion limit at $95\%$ C.L. prospects for a $Z'$ gauge boson coupling predominantly to muons on the $Z'$ coupling-mass plane. The MuSIC reach (red) is compared with a Beam Dump experiment~\cite{Cesarotti:2023sje} that utilizes a $1.5\,\TeV$ muon beam energy and 5\,m thick lead target plate (blue dashed). The gray region denotes the current bounds from BaBar~\cite{Batell:2016ove,BaBar:2016sci} and trident production~\cite{CHARM-II:1990dvf,CCFR:1991lpl,Altmannshofer:2014pba}. The future prospects for the HL-LHC~\cite{Galon:2019owl} (dashed black), FASER$\nu$2~\cite{Ariga:2023fjg} (dashed gray), NA64$\mu$~\cite{Chen:2018vkr} (dotted gray), and $M^3$~\cite{Kahn:2018cqs} (dot-dashed gray) are also shown.}
    \label{fig:Zprime_bounds.pdf}
\end{figure}

\vskip0.5cm

In this Section we compare the potential reach of the MuSIC against Muon-Beam-Dump experiments, which naturally excel in detecting New Physics effects that are specific to muons. 
One such scenario is a new $U(1)'$ vector boson, generically called $Z'$~\cite{He:1990pn,Foot:1990mn,He:1991qd}, which couples dominantly to muons and its mass lies well below the electroweak scale.  This can be implemented by gauging an accidental SM symmetry or by mixing with heavy vector-like leptons,  e.g.~\cite{Kamenik:2017tnu,Delaunay:2020vdb,Krnjaic:2019rsv}.
As a benchmark we consider vector couplings described by the effective interaction Lagrangian
\begin{align}
    \cL_{Z'}^{\rm int} 
    &= 
    -g_{Z'}^{\mu}  \bar{\mu} \gamma_\alpha \mu Z'^\alpha\,.
\end{align}
We note that we expect similar results in the case of an  additional axial-vector or spin-0 coupling.
Finally, it may be that the new physics contribution to the muon anomalous magnetic moment is suppressed due to cancellation at the quantum level~\cite{Balkin:2021rvh}, thus motivating searches for tree-level process as we present here. 

The process of interest is the muon bremsstrahlung $\mu\,\text{Au} \to \mu \,\text{Au}\,Z'$ upon coherent scattering from gold ions ($Z=79$). 
If the momentum transfer $Q^2$ between the muon and the ion is small the electromagnetic scattering through a virtual photon can be coherent over the ion, leading to a $\propto Z^2$-enhancement of the cross section.
The enhancement is reduced as the mass of the emitted particle
is increased, nevertheless, the mass scale at which this happens is larger for the MuSIC due to its larger collisional energy budget compared to a typical Beam Dump setup with a target at rest in the laboratory frame. 

Additionally, the large ion-frame muon energy of $\cO(100\,\TeV)$ leads to highly-boosted particles in the direction of the lepton beam, which is known as \emph{far-backward} region.
After traveling a certain distance $\ell$ the $Z'$ will decay to two muons with branching ratio 1 and probability density
\begin{equation} \label{eq:decay_prob_displaced}
    P(\ell) = \frac{e^{-\ell/L_{Z'}}}{L_{Z'}}\,, \qquad L_{Z'} = \beta \gamma / \Gamma_{Z'}\,,
\end{equation}
where the decay width of the muonphilic $Z'$ is 
\begin{equation} \label{eq:decay_width_Zp}
    \Gamma_{Z'} = (g_Z^{\mu})^2 \frac{m_{Z'}}{12\pi} \left( 1 + \frac{2 m_{\mu}^2}{m_{Z'}^2} \right)\left( 1 - \frac{4 m_{\mu}^2}{m_{Z'}^2} \right)^{1/2}\,.
\end{equation}
The boost factor $\beta \gamma = |\vec{p}|/m_{Z'}$ can be calculated given the distributions of the $Z'$ three-momenta (see Appendix \ref{app:distributions_Zprime} for a few benchmark masses).

Then one can calculate the probability that the $Z'$ decays within the fiducial volume of interest $\cP(\ell_{\rm min}, \ell_{\rm max})$ by integrating \eqref{eq:decay_prob_displaced} from a distance $\ell_{\rm min}$ away from the interaction point to the maximum distance $\ell_{\rm max}$, where a detector that can reconstruct the product particles is located. 
We deploy then a displaced-vertex search strategy, where the expected number of events is given by
\begin{align}
    N_{\rm tot} 
    =& 
    L_I \int d\gamma d\eta \frac{d^2\sigma_{\mu \text{Au} \to \mu \text{Au} Z'}}{d\gamma d\eta} 
    \epsilon_{\rm det}\cP(\ell_{\rm min}, \ell_{\rm max})\,,
\end{align}
where $\gamma$ is the boost factor in the lab frame, $\eta$ is the $Z'$ pseudo-rapidity, and $\epsilon_{\rm det}$ is the detection efficiency of the decay products, which for di-muons we take it to be $86\,\%$~\cite{Bandyopadhyay:2022klg}. 
We choose $\ell_{\rm min}=1\,{\rm mm}$ to reject backgrounds arising from prompt tracks~\cite{Davoudiasl:2023pkq}, and assume that this suffices to conduct a background-free search.
Notice that also the respective studies for Beam Dump experiments work under the background-free assumption.

Due to the large boost factors, in order to capture a significant fraction of the BSM events, a large pseudo-rapidity $\eta$ coverage in the far-backward region is required.
While of course there is no definitive design for the MuSIC facility at the moment, we assume a similar detector infrastructure as the ECCE proposal for the forward region of the EIC~\cite{Adkins:2022jfp,Jentsch} together with a far-backward muon calorimeter~\cite{Acosta:2022ejc}. 
In particular, charged particles with pseudo-rapidities $\eta<6$ will be reconstructed in the EIC within the B0 spectrometer, which will be placed close to the central region of the experiment. 
In our case, we consider a muon spectrometer able to measure muons with reasonable resolution up to $\eta_{\rm max} = -8$. 
In order to achieve the small-angle coverage, the spectrometer must be placed at larger distance downstream of the central region, similarly to the off-momentum detectors in the forward case~\cite{Adkins:2022jfp}. 
Given that a luminosity monitor will be stationed at the EIC in the backward region, taking $ \ell_{\rm max} = 30\,{\rm m}$ we make sure that the spectrometer does not obstruct it.

We compare our MuSIC projection to the Muon Beam Dump experiment proposed in Ref.~\cite{Cesarotti:2023sje}:  A 5\,m thick lead target followed by 10\,m of shielding and a detector far downstream (100\,m). 
The results are obtained using a muon beam of $1.5\,\TeV$ and $N_T=10^{20}$ muons on target. Consequently, this setup enjoys a much larger effective luminosity due to the far larger number density of the target nucleons compared to the MuSIC ion beam, but this feature would not help at higher masses for the produced BSM state, due to more limited kinematics. 

Our results are presented in Fig.~\ref{fig:Zprime_bounds.pdf}. 
Most of the parameter space of the exclusively muonphilic $Z'$ that the MuSIC can probe is not constrained by current experiments ({\it i.e.}  BaBar~\cite{Batell:2016ove,BaBar:2016sci} and trident production~\cite{CHARM-II:1990dvf,CCFR:1991lpl,Altmannshofer:2014pba}) and thus constitutes a suitable benchmark for comparative studies. 
We see that a MuSIC with $1\,\TeV$ muon beam energy provides excellent reach for $Z'$ masses from the di-muon kinematical threshold $2 m_{\mu}$ up to $\sim 20\,\GeV$. 
In comparison the 5\,m thick Beam Dump requires a $1.5\,\TeV$ muon beam energy and a detector at large distances to push the limit to even lower couplings. 
Finally, the HL-LHC is not expected to set any competitive bounds~\cite{Galon:2019owl}, while FASER$\nu$~\cite{Ariga:2023fjg}, NA64$\mu$~\cite{Chen:2018vkr}, and $M^3$~\cite{Kahn:2018cqs} are all future experiments that may provide complementary bounds for $g_{Z'}^{\mu} \lesssim  10^{-4}$ and $m_{Z'} \lesssim \ord{\GeV}$.

\section{Axion-like particles}
\label{sec:ALPs}

\begin{figure}
 \centering
 \includegraphics[width=0.45\linewidth]{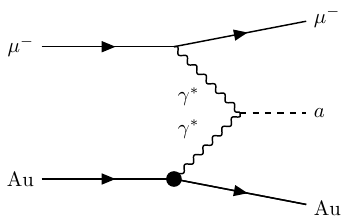}
 \includegraphics[width=0.45\linewidth]{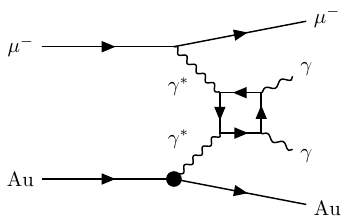}
 \includegraphics[width=0.75\linewidth]{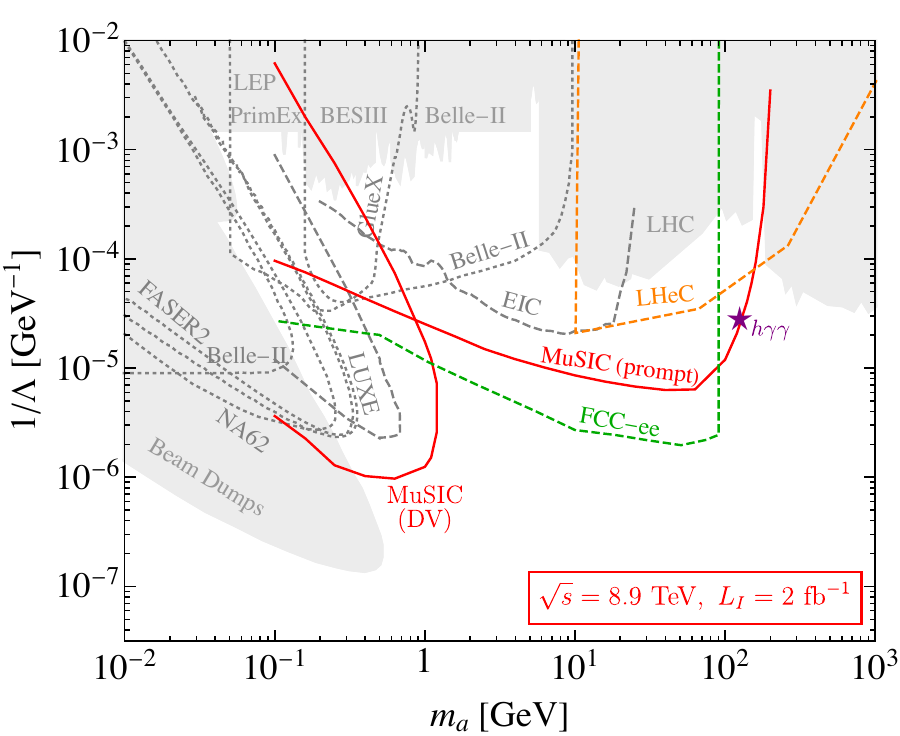}
    \caption{Top: The Feynman diagrams for the ALP production via photon fusion (left) and the main SM background (right). Bottom: The 2~$\sigma$ MuSIC projections of the  ALP-photon coupling from both prompt-decay and displaced-vertex searches as a function of the ALP mass. The current bounds, shaded by gray, are taken from Refs.~\cite{Bjorken:1988as,OPAL:2002vhf,Dobrich:2015jyk,Jaeckel:2015jla,Knapen:2016moh,Bauer:2017ris,Aloni:2019ruo,PrimEx:2010fvg,Belle-II:2020jti,BESIII:2022rzz,Pybus:2023yex}.
    The dotted gray curves denote the FASER2~\cite{Feng:2018pew}, NA62~\cite{Dobrich:2015jyk,Dobrich:2019dxc}, LUXE-NPOD~\cite{Bai:2021gbm}, GlueX~\cite{Aloni:2019ruo,Pybus:2023yex}, Belle-II~\cite{Dolan:2017osp}, FCC-ee~\cite{RebelloTeles:2023uig}, and LHeC~\cite{Yue:2019gbh} projections and the dashed gray curves the EIC ones~\cite{Balkin:2023gya}. The purple star denotes the Higgs-photon coupling value that the MuSIC could additionally probe.}
    \label{fig:ALP_bound} 
\end{figure}
Axion-like particles (ALPs) are pseudoscalar particles that arise naturally in many extensions of the SM. 
The QCD axion provides a solution to the strong $CP$ problem~\cite{Peccei:1977hh,Peccei:1977ur}. 
In this work, we focus on ALPs with masses $m_a$  much larger than the QCD axion mass. Heavier ALPs can still be connected to the QCD axion~\cite{Berezhiani:2000gh,Hook:2014cda,Fukuda:2015ana,Gherghetta:2016fhp,Dimopoulos:2016lvn,Agrawal:2017ksf,Gaillard:2018xgk,Hook:2019qoh,Gherghetta:2020ofz,Csaki:2019vte,Gavela:2023tzu} (even alleviating the axion quality problem~\cite{Agrawal:2017ksf,Hook:2019qoh}), but they can also be motivated in inflation~\cite{Takahashi:2021tff} and dark portal models~\cite{Nomura:2008ru,Dolan:2014ska,Hochberg:2018rjs,Fitzpatrick:2023xks}. 
The ALP $a$ interacts with the SM particles via a coupling suppressed at low energies by a cut-off scale $\Lambda$. 
In this work, we assume that the predominant coupling is to photons,
\begin{equation}
    \cL_a^{\rm int} 
    = 
    -\frac{a}{4\Lambda}F^{\mu\nu}\tilde{F}_{\mu\nu}\, ,
\end{equation} 
where $F_{\mu\nu}$ is the photon field strength tensor with $\tilde{F}_{\mu\nu} = \frac{1}{2} \epsilon_{\mu\nu\alpha\beta}F^{\alpha\beta}$.

Similarly to the EIC case~\cite{Balkin:2023gya}, ALPs can be produced coherently via photon fusion $\mu \,\text{Au} \to \mu \,\text{Au}\,a$ (see upper left diagram in Fig.~\ref{fig:ALP_bound}), followed by the axion decaying into photon pairs with the rate of $\Gamma_{a\to\gamma\gamma} = m_a^3/(64\pi\Lambda^2)$. 
Depending on the mass of the ALP and the coupling, the diphoton searches can be either prompt or displaced. 
We assume a perfect photon detection efficiency~\cite{AbdulKhalek:2021gbh} after satisfying the cuts $E_\gamma > 1.0\,\GeV$ and $\mid\eta_\gamma\mid < 4.0$. 
The main background is due to the irreducible light-by-light scattering~\cite{Bern:2001dg} (see the upper right diagram in Fig.~\ref{fig:ALP_bound}).
In the displaced-vertex search, we take 10\,cm for the displaced diphoton-vertex resolution, and 1\,m for the length of the detector. 
We also assume that the backgrounds are negligible. 

We show the derived MuSIC projections with 400/A\,$\fb^{-1}$ integrated luminosity by the red curves in the lower panel of Fig.~\ref{fig:ALP_bound}. 
The dashed gray curves show the projections of the EIC with 3\,$\fb^{-1}$ integrated luminosity. 
Given that muon beams carry significantly higher energy than the EIC electron beams, the MuSIC can probe ALPs up to masses of approximately 200\,GeV.
Compared to the current bounds from beam-dumps~\cite{Bjorken:1988as,Dobrich:2015jyk}, LEP~\cite{Jaeckel:2015jla}, Belle-II~\cite{Belle-II:2020jti}, BESIII~\cite{BESIII:2022rzz}, PrimEx~\cite{Aloni:2019ruo,PrimEx:2010fvg}, GlueX~\cite{Pybus:2023yex} and ATLAS/CMS~\cite{Knapen:2016moh,Bauer:2017ris}, we see that the MuSIC has the potential to probe unexplored regions of the parameter space.  
 Additionally, the displaced-vertex search at the MuSIC can reach $m_a\sim1\,\GeV$ with $\Lambda\sim 10^6\,\GeV$. The SHiP experiment can potentially set strong bounds on the ALP-photon couplings as seen in Ref.~\cite{Dobrich:2015jyk} based on its original technical proposal~\cite{SHiP:2015vad}.  However, the new version of the experiment is more compact~\cite{Albanese:2878604} and requires a re-analysis, which is not available yet.

We also draw the comparison with the other proposal for the next generation lepton-ion collider, i.e. LHeC, as well as with FCC-ee. 
The dominant production channel is again the photon fusion at both machines. The projected bound from LHeC~\cite{Yue:2019gbh} with $\sqrt{s}=1.3\,\TeV$ and $L_I=1\,\rm ab^{-1}$, is given by the dashed orange curve. 
The bound for Run 1 of FCC-ee~\cite{RebelloTeles:2023uig}, which would operate at the $Z$-pole $\sqrt{s}=91\,\GeV$ with an enormous integrated luminosity $204\,\rm ab^{-1}$, is given by the dashed green curve. 
We observe that the MuSIC outperforms the LHeC in this search, as well as for the contemporaneous Run of FCC-ee for $m_a>m_Z$. However, FCC-ee provides a better reach to lower coupling values. We also note that the LHeC study was performed using protons and if it were to be repeated with ions, we would expect comparable results due to the coherent enhancement.

Before closing this Section, it is worth considering also the Higgs coherent production via the effective SM Higgs-photon coupling
\begin{equation}
    \cL_h^{\rm int} 
    = 
    -\frac{h}{4\Lambda_{h\gamma\gamma}}F^{\mu\nu}F_{\mu\nu}\,,
\end{equation}
where $\Lambda_{h\gamma\gamma} = \pi v_{\rm EW}/(2\alpha I)$, with $I \simeq -1.46$. 
It is straightforward to check that the cross section of the photon fusion process with either scalars or pseudo-scalars are identical under the effective-photon approximation. 
Consequently, the effective Higgs-photon coupling $1/\Lambda_{h\gamma\gamma} \simeq 2.8\times 10^{-5}$~GeV$^{-1}$, which is given in Fig.~\ref{fig:ALP_bound} by the purple star, can be probed at the MuSIC as well.

\section{Heavy sterile neutrinos}
\label{sec:heavy_neutrinos}

\begin{figure}
 \centering
  \includegraphics[width=0.45\linewidth]{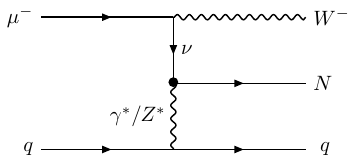}
 \includegraphics[width=0.45\linewidth]{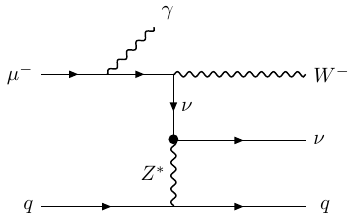}
 \includegraphics[width=0.47\linewidth]{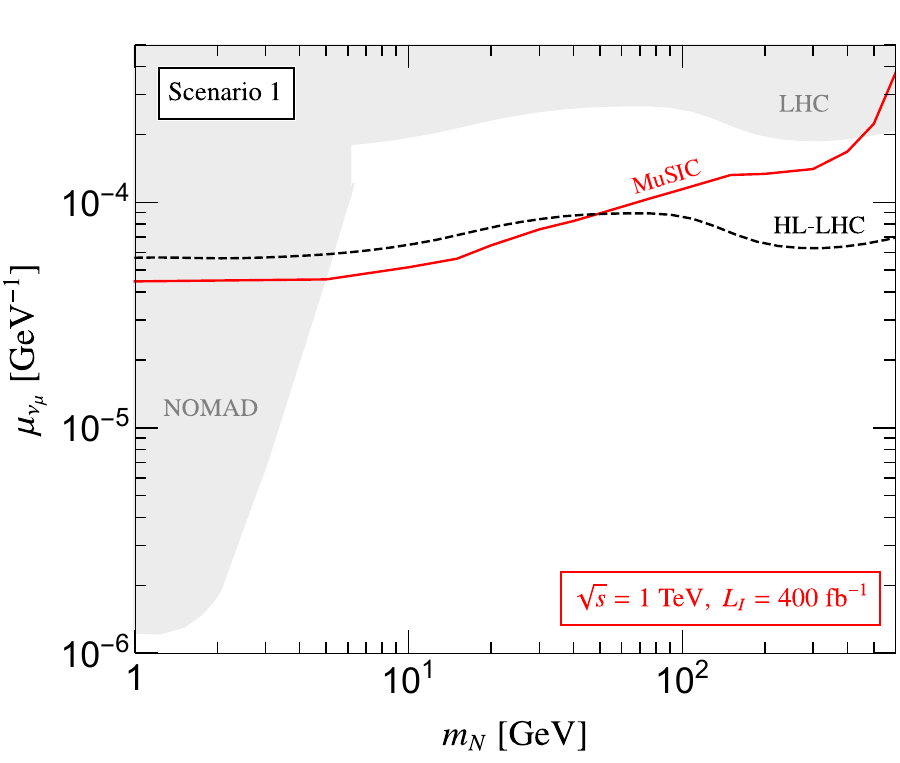}
 \includegraphics[width=0.47\linewidth]{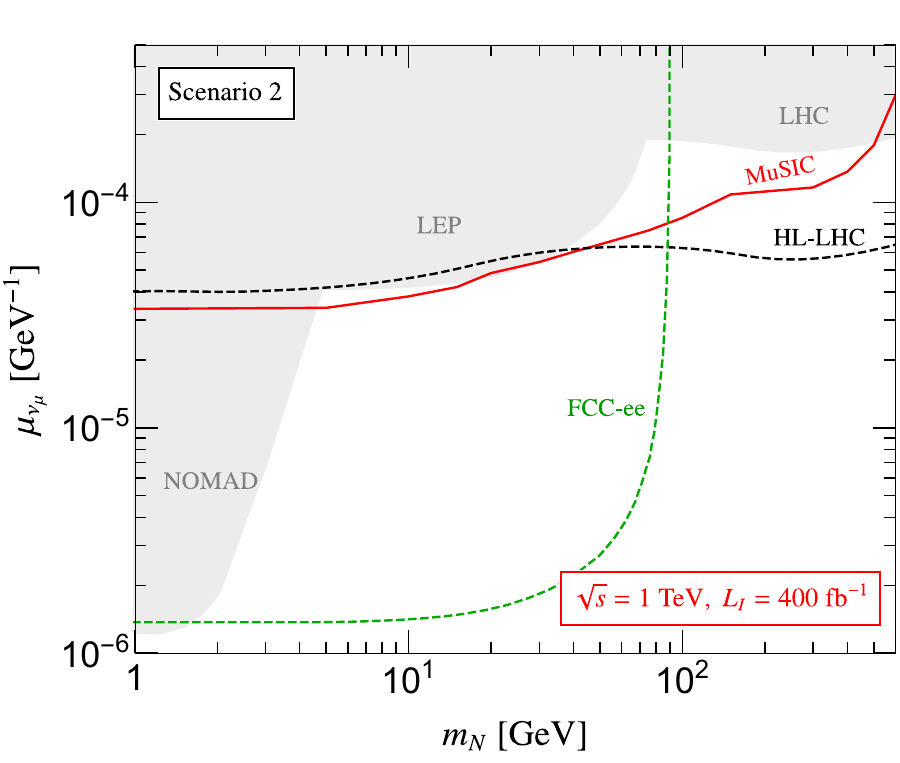}
\caption{Top: The Feynman diagrams for the production of heavy sterile neutrino $N$ via dipole operator in Eq.~(\ref{eq:dim5}) (left) and that of the main SM background (right). Bottom: The 2~$\sigma$ projections of $\mu_{\nu_\mu}$ as a function of the sterile neutrino mass. The red curve shows the MuSIC projection with $400\,\rm fb^{-1}$, whereas the dashed black and green  curves represent the HL-LHC and FCC-ee projections, respectively. The current bounds from NOMAD~\cite{Gninenko:1998pm,Gninenko:1998nn}, LHC~\cite{ATLAS:2017nga,Magill:2018jla}, and LEP~\cite{Magill:2018jla} are shaded by gray.}
\label{fig:dipole_bound}
\end{figure}

Effective operators associated with neutrinos can be probed well at high-energy colliders. 
In this Section, we consider a neutrino up-scattering to a heavy sterile neutrino $N$ via a dipole operator, which can be described by an effective dimension-5 operator 
\begin{equation}
    \label{eq:dim5}
    \cL^{(5)}_{\rm dipole} 
    \supset 
    \frac{1}{2}\mu_\nu\bar\nu F_{\mu\nu}\sigma^{\mu\nu} N\,,
\end{equation}  
where $\mu_\nu$ is the strength of the active to sterile transition neutrino magnetic moment. This operator has been studied both at high-energy~\cite{Magill:2018jla,Ismail:2021dyp,Zhang:2022spf,Ovchynnikov:2023wgg,Beltran:2024twr,Barducci:2024kig} and low-energy experiments~\cite{Coloma:2017ppo,Shoemaker:2018vii,Brdar:2020quo,Schwetz:2020xra,Atkinson:2021rnp,Bolton:2021pey,Barducci:2023hzo,Li:2024gbw}. At sufficient high energy, where the electroweak symmetry is restored, one should consider the relevant dimension-6 $\nu$SMEFT operators
\begin{equation}
\mathcal{L}^{(6)}_{\rm dipole} \supset \frac{c_B}{\Lambda^2}g^\prime B_{\mu\nu}\bar L^\alpha \tilde{H}\sigma^{\mu\nu}N + \frac{c_W}{\Lambda^2}g W^a_{\mu\nu}\bar L^\alpha \sigma^a\tilde{H}\sigma^{\mu\nu}N\,.
\end{equation}  
From those dimension-6 operators, the interaction with $W$ and $Z$ bosons will be generated simultaneously
\begin{equation}
\mathcal{L}_{\rm dipole} \supset \frac{1}{2}\mu_\nu \bar\nu F_{\mu\nu}\sigma^{\mu\nu} N + d_Z \bar\nu Z_{\mu\nu}\sigma^{\mu\nu} N + d_W \bar\ell W^-_{\mu\nu}\sigma^{\mu\nu} N\, ,
\end{equation} 
with the strength
\beqa
\mu_\nu &=& \frac{\sqrt{2}ve}{\Lambda^2}(c_B+c_W)\,,\\
d_Z &=& \frac{ve}{\sqrt{2}\Lambda^2}(c_B\tan\theta_W-c_W\cot\theta_W)\,,\\
d_W &=& \frac{ve}{\Lambda^2}\frac{c_W}{\sin\theta_W}\,.
\eeqa
In the following, we only focus on 2 scenarios,
\begin{enumerate}
    \item only $\mu_\nu \neq 0$\,,
    \item $c_W = 0$ and $d_Z = \frac{1}{2}\mu_\nu \tan\theta_W = \frac{ve}{\sqrt{2}\Lambda^2}c_B \tan\theta_W$\,.
\end{enumerate}
In the first scenario, we only turn on the effective dimension-5 operator in Eq.~(\ref{eq:dim5}),  while we turn off the SU(2) portal at dimension-6 in the second scenario.
At the MuSIC, the heavy sterile neutrino $N$ can be produced through the up-scattering of SM neutrinos $q\mu \to q W N$ via the dipole operators as shown in the upper left panel of Fig.~\ref{fig:dipole_bound}. 
The sterile neutrino further decays into an SM neutrino and a photon. 
The main SM background is shown in the upper right panel of Fig.~\ref{fig:dipole_bound}. 
Depending on the boost of $N$ in the lab frame, the azimuthal angle between the missing energy and the photon, $\Delta\varphi_{\gamma,\rm miss}$, can either be collimated or separated by an angle of $\pi$. 
This characteristic can be very distinct from the background. 
To fully reconstruct the transverse momentum of the SM neutrino from $N$ decay, we require no additional neutrinos produced in the final state. 
Thus, we consider the decay of $W^-$ into two jets. 
The background would include any process that produces $\nu + \gamma + 3$ jets in the final state. 
We simulate the signal and background by using \texttt{MadGraph5\_aMC@NLO}~\cite{Alwall:2011uj} and \texttt{FeynRules}~\cite{Christensen:2008py}. 
There are three di-jet pairs, each forming a di-jet invariant mass. 
We select the candidate mass $M_c$ as the one closest to the value of $M_W$, and then we require $|M_c - M_W| < 5 \,\GeV$ to remove much of the background. 
We further use the information on $\Delta\varphi_{\gamma,\rm miss}$, $\eta_\gamma$, and $p_{T,\gamma}$ to maximize the sensitivities.
We provide the details of the kinematic distributions in Appendix~\ref{app:distributions_N}.

Fig.~\ref{fig:dipole_bound} depicts our results, where again the MuSIC projection is represented by the red curve. 
The LHC bound in the first scenario is derived by rescaling the LHC bound with 36.1 fb$^{-1}$ in the second scenario from Ref.~\cite{Magill:2018jla}, based on the cross section ratios between the process $pp\to\gamma^*\to (N\to \nu_\mu \gamma)\bar\nu_\mu$ and $pp\to \gamma^*/Z^*\to(N\to \nu_\mu \gamma)\bar\nu_\mu$.
The cross section ratios are calculated with \texttt{MadGraph5} after applying the cuts $p_{T,\gamma}, E_{T, \rm miss} > 150$~GeV. 
We see that thanks to the energetic muon and highly boosted proton beams, the MuSIC will be able to probe heavy sterile neutrinos up to $500\,\GeV$, extending significantly beyond the current bounds~\cite{Gninenko:1998pm,Gninenko:1998nn,ATLAS:2017nga,Magill:2018jla} in the first scenario. The MuSIC is comparable to the LEP1 with 200 pb$^{-1}$ ~\cite{Magill:2018jla} ($\sim$ 4 million $Z$ bosons~\cite{AMANN200217}) for masses below $20\,\GeV$ in the second scenario.
The HL-LHC and FCC-ee projections are estimated by rescaling the LHC and LEP1 bound using $3\,\rm ab^{-1}$ and $204\,\rm ab^{-1}$ integrated luminosity, respectively. 
Evidently, the HL-LHC will only be capable of surpassing MuSIC's performance for masses above approximately $50\,\GeV$. On the other hand, FCC-ee has the potential to place the most stringent bound from $\mathcal{O}(1)$~GeV to $m_Z$ in the second scenario. 

\section{Conclusions}
\label{sec:conclusions}

\begin{table}[t]
\begin{center}
\begin{tabular}{ | c | c | c | }
\hline
\textbf{BSM candidate} & \textbf{Search} & \textbf{Features} \\
\hline\hline
LFV $U_1$ leptoquarks & prompt  & valence lepton \\
\hline
Muonphilic $Z'$ & displaced & \makecell{far-backward detector, \\ coherent scattering, \\ valence lepton} \\
\hline
Axion-like particles & \makecell{prompt \& displaced} & coherent scattering \\
\hline
Sterile Neutrinos & prompt  & mono-neutrino \\
\hline
\end{tabular}
\caption{Summary of the New Physics scenarios analyzed and the respective key advantages of the MuSIC.} \label{tbl:summary}
\end{center} 
\end{table}

The MuSIC is a next generation lepton-ion collider with dual appeal: it promises to establish a new QCD frontier while also facilitating the development of a muon synchrotron technologies, paving the path towards a multi-TeV MuC. 
In this work, we have further emphasized the science potential of the MuSIC by conducting a first study of its New Physics opportunities. 
In particular, we scrutinized four motivated scenarios with explicit BSM mediators both below and above the energy threshold of $\sqrt{s}=1\,\rm TeV$. 
In all cases, we found that the MuSIC either outperforms the competing future experiments or complements them. 
This conclusion holds, even when the MuSIC is compared against experiments with much higher effective luminosity, \eg Muon Beam Dumps, or center-of-mass energy, \eg the HL-LHC. 
The key behind the exceptional discovery reach of the MuSIC is that it simultaneously leverages various unique features of both muon beams as well as the accelerator and detector systems of the EIC. 
In Table~\ref{tbl:summary}, we summarize the proposed New Physics searches, together with MuSIC's experimental strengths that we have identified as crucial for the success of each search. 

We stress that our work is by no means an exhaustive analysis of all relevant BSM scenarios and that we have merely scratched the surface of the MuSIC's significant potential. 
First of all, one could study popular variations of the models we have considered. To name a few examples, one can consider other types of leptoquarks, LFV $Z'$ gauge vector bosons, ALPs coupling predominantly to other SM particles, as well as different types of fundamental particles or interactions altogether, \eg scalars, dark non-Abelian gauge interactions, and higher-dimensional operators. 

Furthermore, one may draw inspiration from the numerous New Physics searches discussed for LHeC in the literature, \eg modified SM couplings~\cite{Senol:2012fc,Biswal:2014oaa,Cakir:2014swa,Kumar:2015kca,Sun:2016kek,Coleppa:2017rgb}, extensions of the scalar sector~\cite{Bernaciak:2014pna,Das:2015kea,Das:2016eob,Curtin:2017bxr,Sun:2017mue,Azuelos:2017dqw,Das:2018vuk,DelleRose:2018ndz}, low-energy supersymmetry~\cite{Kuday:2013cxa,Li:2013swh,Zhang:2014qkh,Azuelos:2019bwg}, sterile heavy neutrinos~\cite{Antusch:2016ejd,Lindner:2016lxq,Antusch:2019eiz,Das:2018usr}, fermion triplets~\cite{Jana:2020qzn,Das:2020gnt}, and dark photons~\cite{DOnofrio:2019dcp}. 
In those proposals, the investigated New Physics effects often do not depend strongly on the flavor of the valence lepton, and one would thus expect equivalent results for the MuSIC. 
However, even though the center-of-mass energy of the LHeC and MuSIC are in the same ballpark, the individual beam energies, the leptonic PDFs, the backgrounds, and the optimal search strategies may all fundamentally differ. 
Indeed, given the prospects of the MuSIC in the examples we analyzed, it would be interesting to revisit other cases discussed in the context of similar experiments as well as to envision new opportunities. 
The MuSIC is poised to orchestrate groundbreaking discoveries, acting as a crescendo to the current research in particle and nuclear physics.

\vskip0.5cm

Digital data related to this work are available at arXiv.org with the corresponding submission material.

\begin{acknowledgments}
We thank Darin Acosta, Reuven Balkin, Cari Cesarotti, Abhay Deshpande, Wei Li, Ben Ohayon, and Jesse Thaler for fruitful discussions and feedback. 
The work of H.D. and H.L. is supported by the U.S. Department of Energy under Grant Contract DE-SC0012704.  S.T. is supported by the U.S. Department of Energy (DOE) Office of High Energy Physics under Grant Contract No. DE-SC0012567, and by the DOE QuantISED program through the theory consortium “Intersections of QIS and Theoretical Particle Physics” at Fermilab (FNAL 20-17). S.T. is additionally supported by the Swiss National Science Foundation - project n.~P500PT\_203156. 
The work of Y.S. is supported by grants from the NSF-BSF (grant No. 2021800) and the ISF (grant No. 597/24).
\end{acknowledgments}

\appendix

\section{\boldmath Kinematical distributions for $\mu \text{Au} \to \mu \text{Au} Z'$ }
\label{app:distributions_Zprime}

\begin{figure}
 \centering
 \includegraphics[width=0.45\linewidth]{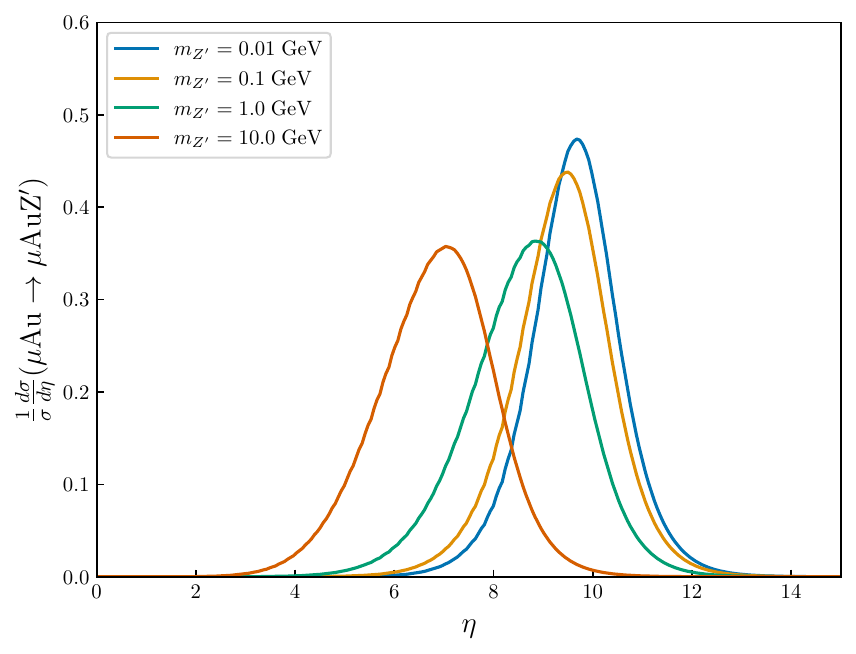}
  \includegraphics[width=0.45\linewidth]{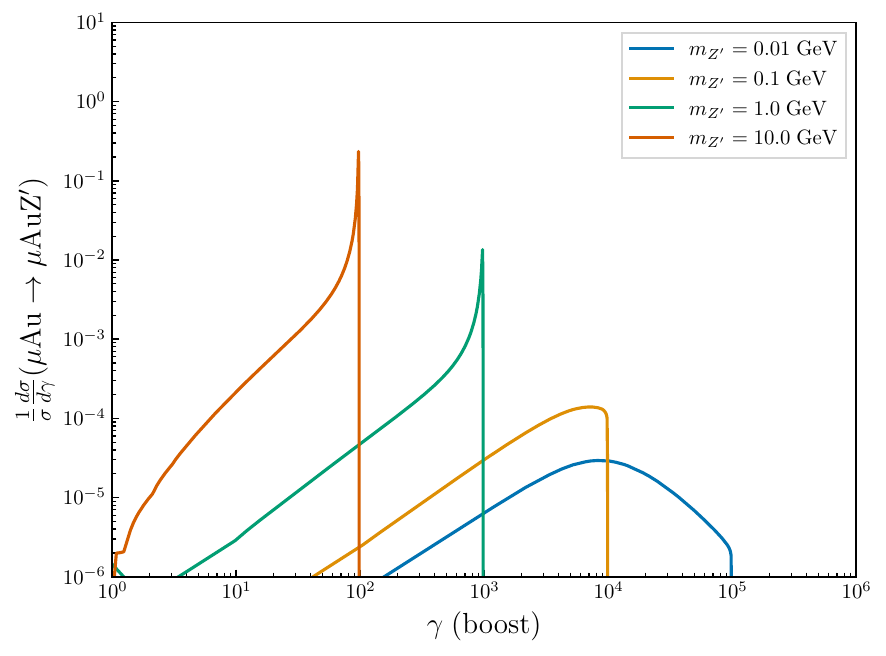}
    \caption{Normalized distributions of the muonphilic $Z'$ boson pseudo-rapidity (left) and boost factor (right) for different $Z'$ masses produced via muon bremsstrahlung $\mu \text{Au} \to \mu \text{Au} Z'$ in the MuSIC.}
    \label{fig:kinematic_distributions}
\end{figure}

The kinematical distributions of the differential cross sections with respect to $\gamma$ and $\eta$ are shown in Fig.~\ref{fig:kinematic_distributions} for various benchmarks of the $Z'$ masses. As is evident, relatively light $Z'$ bosons are produced with large boosts in the direction of the high-energy muon beam. The $\eta$ distributions peak in the region $6\lesssim \eta \lesssim 10$, which justifies the coverage we proposed above in the backward region. 

\FloatBarrier

\section{\boldmath Kinematical distributions for heavy sterile neutrinos production}
\label{app:distributions_N}
\begin{figure}
 \centering
 \includegraphics[width=0.45\linewidth]{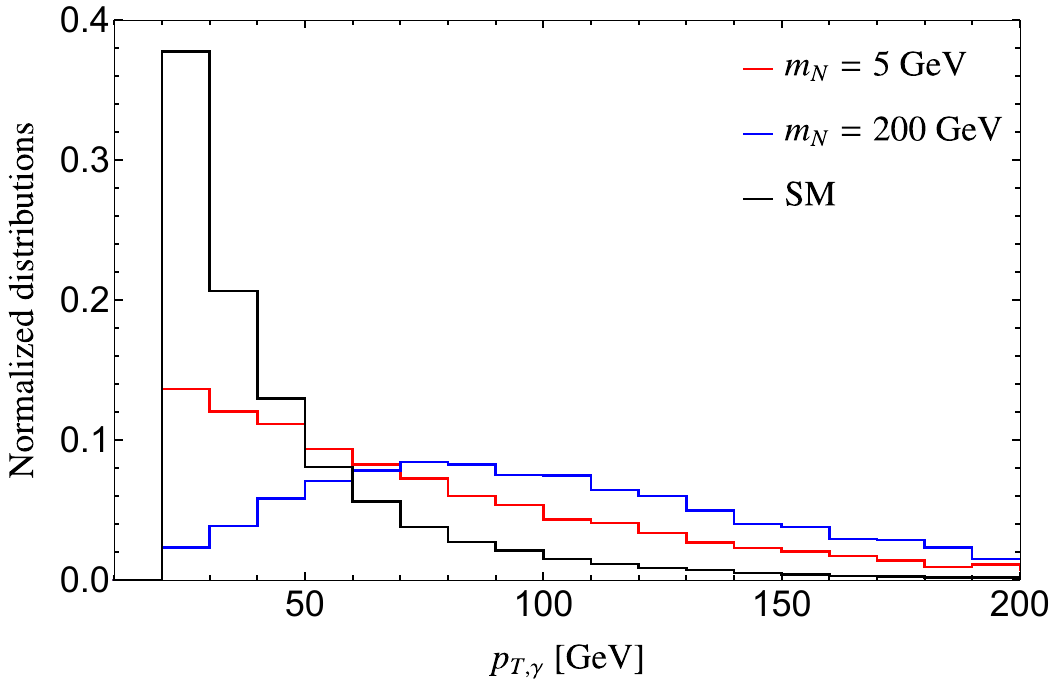}
 \includegraphics[width=0.45\linewidth]{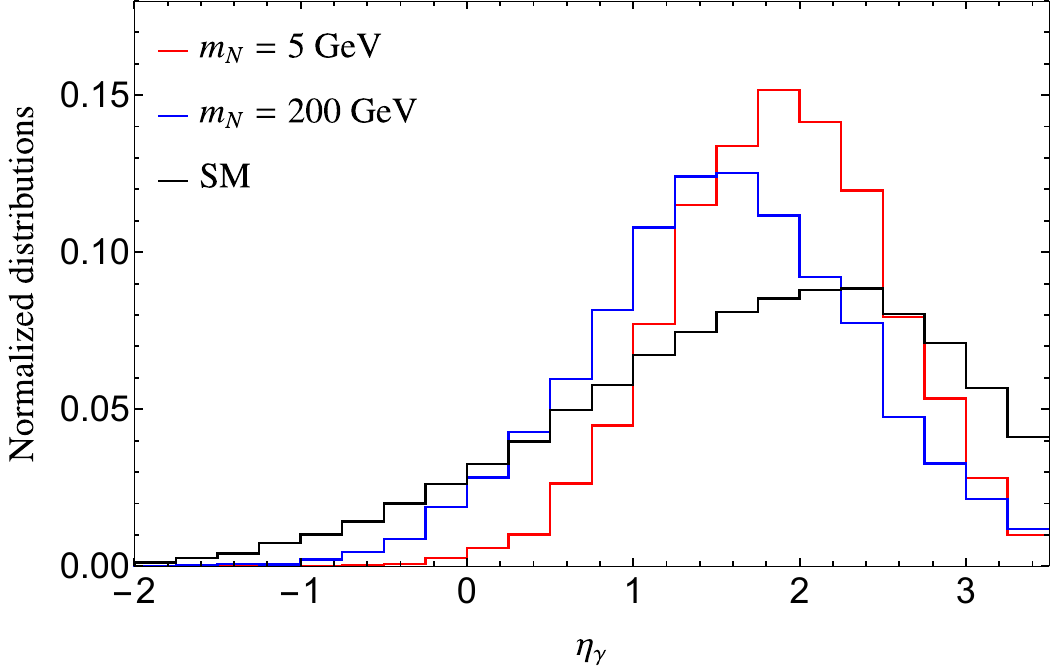}
 \includegraphics[width=0.45\linewidth]{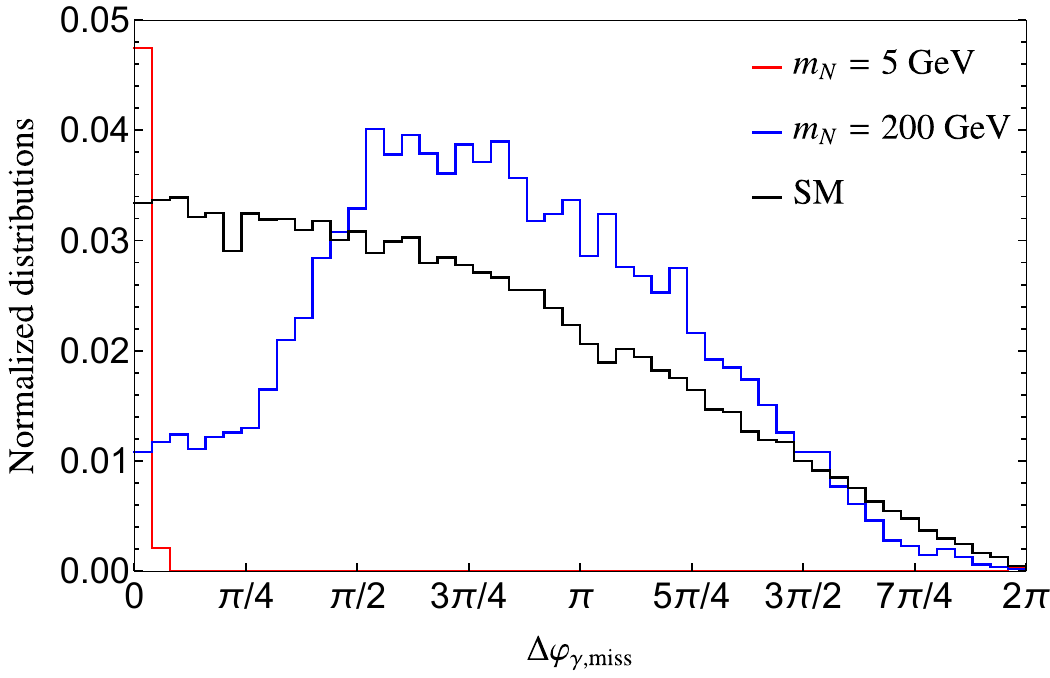}
 \includegraphics[width=0.45\linewidth]{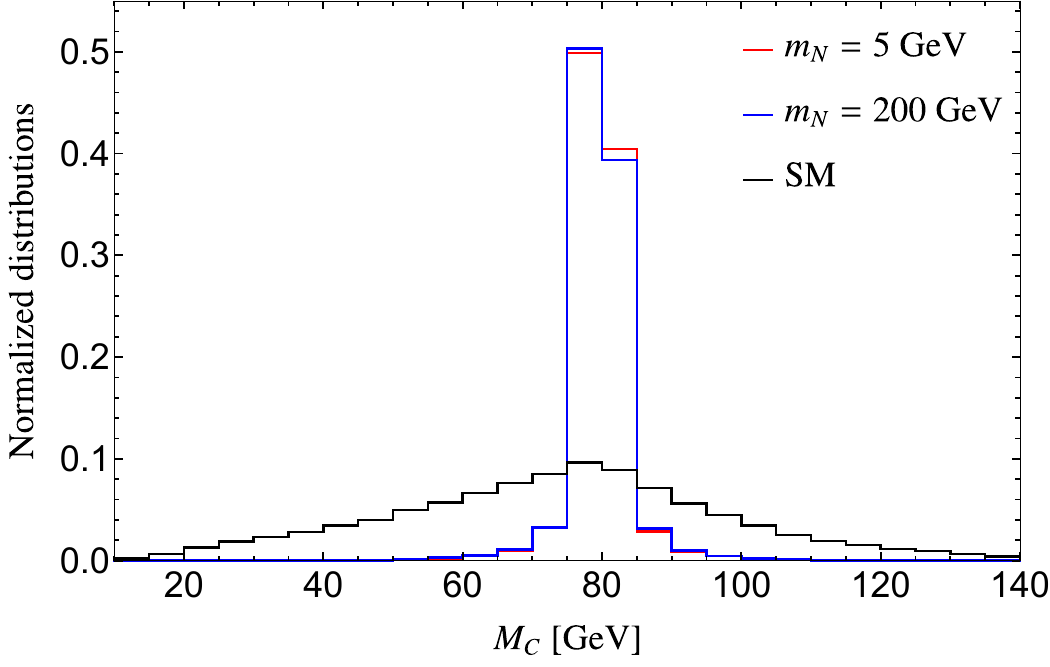}
    \caption{Unity-normalized distributions of the photon transverse momentum (top left), photon pseudorapidity (top right), azimuthal
angle between the missing energy and the photo (bottom left), and the $W$ candidate mass (bottom right). The red~(blue) curves show the results for the signal with $m_N = 5~(200)$~GeV, and the SM distributions are represented by the black curves. The red curve in the bottom left panel is normalized to 0.05 for convenience.}
    \label{fig:kinematic_N}
\end{figure}

We present some important kinematic distributions related to the heavy sterile neutrino $N$ production in Fig.~\ref{fig:kinematic_N}. The normalized distributions of the transverse momentum $p_{T,\gamma}$ and pseudorapidity $\eta_\gamma$ of the final-state photon from $N$ decay are shown in the top left and top right panels of Fig.~\ref{fig:kinematic_N}, respectively. As the mass $m_N$ increases, the photon $p_T$ increases as well and becomes less forward compared to the SM background process. The distributions of azimuthal
angle between the missing energy and the photon $\Delta\varphi_{\gamma,\rm miss}$ are shown in the bottom left panel of Fig.~\ref{fig:kinematic_N}. We see that $\Delta\varphi_{\gamma,\rm miss}$ is close to zero for light $N$. In contrast, for heavy $N$, $\Delta\varphi_{\gamma,\rm miss}$ has a broader distribution and peaks around $3\pi/4$.
The candidate $W$ boson mass $M_c$ for the signal is close to the true value $M_W$ as shown by the bottom right panel of Fig.~\ref{fig:kinematic_N}.

\bibliographystyle{JHEP}
\bibliography{biblio}

\end{document}